\documentclass[showpacs,preprintnumbers,amsmath,amssymb]{revtex4}
\usepackage{graphicx}
\usepackage{dcolumn}
\makeatletter
\input{epsf}
\begin{document}
\title{Charge and Statistics of Quasiparticles in Fractional Quantum Hall Effect }
\author{B. Basu}
 \email{banasri@isical.ac.in}

\author{P. Bandyopadhyay}
 \email{pratul@isical.ac.in}
 \author{S. Dhar}
\email{sarmi_30@rediffmail.com}
\affiliation{Physics and Applied Mathematics Unit\\
 Indian Statistical Institute\\
 Kolkata-700108 }
\begin{abstract}
We have studied here the charge and statistics of quasiparticle
excitations in FQH states on the basis of the Berry phase approach
incorporating the fact that even number of flux quanta can be gauged
away when the Berry phase is removed to the dynamical phase. It is
observed that the charge $q$ and statistical parameter $\theta$ of a
quasiparticle at filling factor $\nu=\frac{n}{2pn+1}$ are given by
$q=(\frac{n}{2pn+1})e$ and $\theta=\frac{n}{2pn+1}$, with the fact
that the charge of the quasihole is opposite to that of the
quasielectron. Using Laughlin wave function for quasiparticles,
numerical studies have been done following the work of  Kj{\o}nsberg
and Myrheim \cite{KM} for FQH states at $\nu=1/3$ and it is pointed
out that as in case of quasiholes, the statistics parameter can be
well defined for quasielectrons having the value $\theta=1/3$.
\end{abstract}
\maketitle

\section{Introduction}

In an earlier work  Kj{\o}nsberg and Myrheim \cite{KM} have reported
that when one uses Laughlin wave function for fractional quantum
Hall (FQH) states, the corresponding quasielectron (QE) excitations
are found to have charge and statistics different from those of
quasiholes (QH). In fact their numerical studies show that for
system sizes upto 200 electrons, quasiholes have the charge and
statistics parameter $e/3$ and $1/3$ respectively. However for
quasielectrons (QE) the trends regarding the charge show a slow
convergence towards the expected value $-e/3$ and with a finite size
correction for $N$ electrons approximately $-0.13 e/N$, the
statistics parameter $\theta$ has no defined value but might
probably converge to $1/3$ in the thermodynamic limit. On the other
hand Kj{\o}nsberg and Leinaas \cite{KL} have shown that the
composite fermion (CF) picture leads to a definite statistics
parameter for a CFQP. Jeon et.al \cite{J,J1} have concluded that the
composite fermion (CF) picture gives rise to the CFQP charge and
statistics parameter given by $e^*=-\frac{e}{2pn+1}$ and
$\theta=\frac{2p}{2pn+1}$ for the filling factor
$\nu=\frac{n}{2pn+1}$. However numerical studies \cite{J,J1} using
CF wave function show that at filling factor $\nu=1/3$ and $2/5$ the
statistics parameter $\theta$ is found to be $-2/3$ and $-2/5$
respectively. Though the numerical values agree with the theoretical
predictions on the basis of CF model, it has opposite sign. The sign
discrepancy would cast doubt on the fundamental interpretation of
the CF physics. To cope with this problem these authors have
suggested that the insertion of one composite fermion quasiparticle
(CFQP) at a certain position is slightly perturbed by inserting
another CFQP inside the loop which pushes the other very slightly
outward.

From the above findings it appears that any of the models of quantum
Hall fluid seem to satisfy the observations in an unambiguous way.
Here we shall try to show that Laughlin wave function appears to
lead to the charge and statistics parameter for QE compatible with
QH when we take into account that the Berry phase associated with an
electron having even number of flux quanta attached to it can be
removed to the dynamical phase and the residual Berry phase of the
QE is computed considering the reduced magnetid field. Indeed this
helps to improve the result obtained by  Kj{\o}nsberg and Myrheim
\cite{KM} at $\nu=1/3$.

In sec.II we shall briefly sketch the model of QHE on the basis of
Berry phase approach \cite{b1,b2} incorporating the fact that even
number flux quanta can be gauged away when the phase is removed to
the dynamical phase. In sec.III we shall analyze Laughlin wave
functions for QE and QH incorporating this aspect of Berry phase. In
sec.IV we shall reproduce the numerical result of Kj{\o}nsberg and
Myrheim \cite{KM} at $\nu=1/3$ taking into account this
modification.

\section{Fractional Quantum Hall Effect and Berry Phase}
In some earlier papers \cite{b1,b2} we have analyzed the sequence of
quantum Hall states from the viewpoint of chiral anomaly and Berry
phase. In our approach , we have considered the spherical geometry
which was first used by Haldane \cite{hal1} where the electrons are
confined on the surface of a sphere of large radius $R$ with a
magnetic monopole of strength $\mu$ at the centre. In this geometry,
the single electron is represented as a spin $S$, the orientation of
which indicates the point on the sphere about which the state is
localized. The angular momentum shells of the spherical geometry are
the analogy of the Landau levels (LL) of the planar geometry. The
angular momentum relation here is given by

\begin{equation}
{\bf J} ~ = ~ {\bf r} \times {\bf p} - \mu {\bf \hat{r}} , ~ \mu
~=~ 0 , ~\pm ~1 / 2 , ~\pm ~1 , ~\pm ~ 3 / 2 \cdots.
\end{equation}

The lowest Landau level (LLL) is given by the $L=|\mu|$ shell.

It may be remarked here that as $|\mu|$ here corresponds to the
monopole strength we can relate this with the Berry phase. Indeed
$\mu=1/2$ corresponds to one flux quantum  and the flux through the
sphere when there is a monopole of strength $\mu$ at the centre is
$2\mu$. The Berry phase of a fermion of charge $q$ when it moves in
a closed path is given by $e^{i\phi_B}$ with $\phi_B=2\pi q N$, here
$N$ is the number of flux quanta enclosed by the loop traversed by
the particle.

If $\mu$ is an integer, we can have a relation of the form
\begin{equation}\label{prime}
{\bf J} ~=~ {\bf r} \times {\bf p} - \mu {\bf \hat{r}} ~=~ {\bf
r}^{~\prime} \times {\bf p}^{~\prime}
\end{equation}
which indicates that the Berry phase associated with integer $\mu$ may be
unitarily removed to the dynamical phase.
 This suggests that the attachment of
$2p$ vortices(magnetic flux lines) with $p$ an integer
 to an electron effectively leads to the removal of
Berry phase to the dynamical phase. So, FQH states with  $2
\mu_{eff} = 2p + 1$ can be viewed as if one vortex is attached to
the  electron. From the Dirac quantization condition $e \mu_{eff} =
\frac{1}{2} $, we can identify the filling factor $\nu$ with
$\frac{1}{2 \mu_{eff}}$ corresponding to the charge of the particle
given by $\nu e$. For a higher Landau level we can consider the
Dirac quantization condition $e \mu_{eff} = \frac{1}{2} n$, with $n$
being a vortex of strength $2 \ell + 1$. This can generate FQH
states having the filling factor of the form $\frac{n}{2 \mu_{eff}}$
where both $n$ and $2 \mu_{eff}$ are odd integers. Indeed, we can
write the filling factor as \cite{b1,b2}

\begin{equation}
\nu ~=~\frac{n}{2 \mu_{eff}} ~=~ \frac{1}{\frac{2 \mu_{eff}~ \mp
1}{n} \pm \frac{1}{n}} ~=~ \frac{n}{2pn \pm 1} \label{e1}
\end{equation}
 where $2 \mu_{eff} \mp 1$ is an even integer
given by $2pn$. In this scheme, the FQH states with $\nu$ having
the form
\begin{equation}
\nu=\frac{n^{~\prime}}{2pn^{~\prime} \pm 1} \label{e2}
\end{equation}
where $n^\prime$ an even integer can be generated through
particle-hole conjugate states
\begin{equation}
\nu ~=~ 1 - \frac{n}{2pn \pm 1} ~=~ \frac{n (2p - 1) \pm 1}{2pn
\pm 1}=\frac{n^{~\prime}}{2pn^{~\prime} \pm 1} \label{e3}
\end{equation}

These FQH states having even numerator and odd denominator filling
factors can be considered as particle-hole conjugate states in this
scheme.

The segregation of even number of vortices helps us to consider the
removal of the corresponding phase factor to the dynamical phase. To
see this explicitly, we take resort to the planar geometry when the
ground state wave function is given by Laughlin wave function. The
ground state wave function for an $N$-particle system at the filling
factor $\nu=\frac{1}{m}=\frac{1}{2p+1}$ is given by
\begin{equation}
    \psi_m= \prod_{i<j} (z_i^* - z _j^*)^m e^{-\frac{1}{2}\sum |z_i|^2}
\end{equation}
 where $z=\frac{x+i y}{\sqrt{2}l}$, $l$ being the magnetic length $l=\frac{1}{\sqrt{e B}}$
$ (\hbar=c=1)$. Now segregating the even number of vortices we consider the phase of the
Jastrow factor $\prod_{i<j}(z_i^* - z _j^*)^{2p}$. We can display the phases
due to the Jastrow factor by writing
\begin{equation}
\prod_{i<k}(z_i^* - z _k^*)^{2p}    =e^{i2p \sum_{j<k} \phi_{j k}}
\prod_{j<k}|z_j-z_k|^{2p}
\end{equation}
where
\begin{equation}
    \phi_{j k}=i~ ln \frac{z_j^*-z_k^*}{|z_j-z_k|}
\end{equation}

This effectively leads to the many body wave function of the electrons $\psi_e$
to the transformed wave function through the relation
\begin{equation}\label{tr1}
    \phi_{tr} \{\vec{r}_i\}=\phi_{e} \{\vec{r}_i\} \prod_{i<j}
    \displaystyle{\left(\frac{z_j^*-z_k^*}{|z_j-z_k|}\right)^{2p}}
\end{equation}
The unitary transformation (\ref{tr1}) of the wave function,
which may be described as a singular gauge transformation requires a
corresponding transformation of the Hamiltonian. The Schr\"{o}dinger
equation can now be written as \cite{halp1}
\begin{equation}\begin{array}{lcl}
    \displaystyle{\left[\frac{1}{2m_b} \sum_i (p_i+ e {\bf A}
    ({\bf r}_i)-e {\bf a} ({\bf r}_i))^2+V \right] \prod_{j<k}|z_j-z_k|^{2p}
    (z_i^*-z_j^*) e^{-\frac{1}{2}\sum |z_i|^2}}&&\\
~~~~~~~~~~~~~~~~~~~~~~~~~~~~~~~~~~~~~    =E
\prod_{j<k}|z_j-z_k|^{2p} (z_i^*-z_j^*) e^{-\frac{1}{2}\sum |z_i|^2} &&\\
\end{array}
\end{equation}
where $m_b$ is the electron band mass. It is noted that here we have
introduced the additional vector potential ${\bf a} ({\bf r}_i)$
that simulates the effect of the phases of the Jastrow factor. In
fact we have

\begin{equation}
    {\bf a} ({\bf r}_i)=\frac{2p}{2\pi} \phi_o \sum^\prime_i { \vec{\nabla}}_i \phi_{ij}
\end{equation}
where the prime denotes the condition $i\neq j$. Here $\phi_0$ is the unit
magnetic flux   quanta given by $\phi_0 = \frac{h c}{e}$. The corresponding
magnetic field is
\begin{equation}
    \vec{b}_i = 2p ~\phi_0 ~\hat{z} \sum^\prime_i \delta^2 ({\bf r}_i-{\bf r}_j)
\end{equation}

Thus the phase of the Jastrow factor is equivalent to each electron seeing a
flux tube of $2p\phi_0$ on other electron. In a translationally invariant  system ,
the mean field Hamiltonian $H_0$ may be obtained by replacing $\vec{b}({\bf r})$
by the mean value
\begin{equation}
    <\vec{b}>=2\pi ~2p ~n_e
\end{equation}
where $n_e$ is the average electron density and by ignoring the
potential energy $V$, we may now write
\begin{equation}
    H_0 =\displaystyle{\frac{1}{2m_b} \int \psi^\dag ({\bf r})
    [-i\nabla + A^* ({\bf r}) ]^2 \psi ({\bf r}) d^2 r}
\end{equation}
where $\bigtriangledown \times A^*=B^*=B-2\pi ~2p ~n_e$.

 Thus the effect of the removal of the Berry phase associated with
 even number of flux quanta to the dynamical phase is transcribed by
 the reduction of the magnetic field $B$ to $B^*$ with
 \begin{equation}
    |B^*|= \frac{B}{2pn+1}
 \end{equation}
 The Berry phase acquired by an electron when it traverses a closed
 path encircling an area $A$ in relation to the reduced field $B^*$
 is given by
 \begin{equation}\label{b*}
    \phi^* = 2\pi \frac{B^* A}{\phi_0}
 \end{equation}

 For a quantum Hall state with filling factor
 $\nu=\frac{n}{2mn+1}=\frac{1}{2m+1/n}$,
it appears that when even number of flux quanta is gauged away, the
electron is attached with a magnetic flux quantum having strength
$1/n$ and so we will have the phase in the reduced field $B^*$
\begin{equation}\label{b*1}
    \phi^* = 2\pi \frac{nB^* A}{\phi_0}
 \end{equation}

 In the field $B$, this gives rise to
\begin{equation}\label{b*2}
\begin{array}{lcl}
    \phi &=& 2\pi \displaystyle{\frac{nB A}{(2pn+1)\phi_0}}\\
    &&\\
    &=& 2\pi \displaystyle{\frac{n}{(2pn+1)}N}\\
\end{array}
 \end{equation}
 where $N$ is the number of flux quanta enclosed by the path. So
 when a particle having the total number of flux quanta $|2p+1/n|$
 in the field $B$ traverses a closed path encircling $N$ number of
 such particles, the Berry phase is given by
 \begin{equation}\label{2p}
    \phi=2\pi \displaystyle{\frac{n}{(2pn+1)}N}
 \end{equation}

It is noted that in our classification scheme we have identified the
filling factor $\nu$ with the local charge of the particle $q=\nu$
(in unit of e) and so we can define the Berry phase by
\begin{equation}
\phi=2\pi qN
\end{equation}

Evidently this is analogous to the Berry phase acquired by an
electron when it traverses a closed path encircling $N$ number of
electrons but is different from that of a composite fermion in the
CF picture. From eqn.(\ref{2p}) it is observed that when a particle
attached with $2p+\frac{1}{n}$ flux quanta moves about another one
traversing half circle the phase is $\phi=\pi \frac{n}{2pn+1}$ so
that the statistics parameter $\theta=\frac{n}{2pn+1}$.

\section{Charge and Statistics of the Quasielectrons and Quasiholes}
The Laughlin wave function is characterized by the fact that for one
quasihole (QH) at the position $z_0$ it is given by
\begin{equation}\label{ee1}
    \psi^{1qh}_{z_0}=\psi_0~ \triangle^{*m}
    \prod^{N}_{i=1}\left(z^*_i-z^*_0 \right)
\end{equation}
where
\begin{equation}
\psi_0 =e^{-\frac{1}{2}\sum^N_{i=1}|z_i|^2}\nonumber
\end{equation}
$$\triangle=\prod_{j<k} (z_j-z_k)$$
and $m=1/\nu$, $\nu$ being the filling factor. It is noted that $z$
is here given by $\displaystyle{\frac{x+iy}{\sqrt{2}l}}$, $l$ being
the magnetic length. For a quasielectron (QE) at $z_0$ the
corresponding wave function is given by
\begin{equation}\label{ee2}
    \psi^{1qe}_{z_0}=\psi_0~
    \prod^{N}_{i=1}\left(\partial_{z^*_i}-z_0 \right)\triangle^{*m}
\end{equation}

To study the statistics parameter we need to compute the phase
related to the motion of one QE(QH) around another QE(QH). So, we
need the construction of two QHs and QEs wave functions. For two QHs
at the positions $z_a$ and $z_b$, the wave function is
\begin{equation}\label{ee3}
    \psi^{2qh}_{z_a,z_b}=\psi_0~ \triangle^{*m}
    \prod^{N}_{i=1}\left(z^*_i-z^*_a \right)\left(z^*_i-z^*_b \right)
\end{equation}

Similarly for the QEs at $z_a$ and $z_b$, the wave function is given
by
\begin{equation}\label{e4}
    \psi^{2qe}_{z_a,z_b}=\psi_0~
    (\prod^{N}_{i=1}\left(\partial_{z^*_i}-z_a \right)
    \left(\partial_{z^*_i}-z_b \right))\triangle^{*m}
\end{equation}

 Now following Kj{\o}nsberg and Myrheim \cite{KM} we consider a quasiparticle
excitation (QH or QE) of charge $q$ localized at the position $z_0$
and is described by the normalized wave function
$<z_1,.....z_N|z_0>$. If this quasiparticle (QP) is moved around a
closed path there arises the Berry phase. If the path is a circle
around the origin parameterized by $z_0=r~ e^{i\phi}$ with $\phi$
running from $0$ to $2\pi$, the Berry connection is then defined by
\begin{equation}
    \frac{d \beta_1}{d\phi}=i<z_0|\partial_{\phi}|z_0>
\end{equation}

 The charge $q$ of the quasiparticle is determined by setting the
 Berry phase equal to the Aharonov-Bohm phase corresponding to the
 same path. Evidently in a finite system the charge defined in this
 way will depend on the distance $r$ from the origin but it gives a
 reasonable value when the circles are well within the droplet. If
 the quasiparticle is described by a normalized $N$- particle state
 of the form
 \begin{equation}\label{i1}
    |z_0>=\frac{1}{\sqrt{I_1}}~\sum^{\alpha}_{l=0} z_0^l~ a_l |l>
\end{equation}
where $a_l$ are expansion coefficients, $|l>$ are orthonormal basis
states and $I_1$ is the normalization factor
\begin{equation}
I_1=\sum^{\alpha}_{l=0} ~r^{2l} ~|a_l|^2
\end{equation}

The expression for charge is given by \cite{KM}
\begin{equation}
    \frac{q}{e}~=~\frac{1}{r^2}~\frac{d
    \beta_1}{d\phi}~=~-\frac{d}{dr^2}~\ln I_1
\end{equation}

When there are two quasiparticle excitations simultaneously located
symmetrically about the origin at the positions $\pm z_0$ with the
parametrization $z_0=r ~e^{i\phi}$ we can let $\phi$ run from $0$ to
$\pi$ which describes the interchange of two quasiparticles. If the
quasiparticle state is described by a state analogous to (\ref{i1})
\begin{equation}
|z_0,-z_0>=\frac{1}{\sqrt{I_2}}~\sum^{\alpha}_{l=0} z_0^l~ b_l |l>
\end{equation}
then a Berry phase connection $\frac{d\beta_2}{d\phi}$ corresponding
this interchange is given by
\begin{equation}
\frac{d\beta_2}{d\phi}=i<z_0,-z_0|\partial_{\phi}|z_0,-z_0>
\end{equation}

We can now define the statistics parameter by subtracting the single
particle contribution due to the magnetic field
\begin{eqnarray}
  -\theta &=& \frac{1}{\pi} (\beta_2(\pi)-2 \beta_1(\pi)) \nonumber \\
   &=& \frac{d}{d\phi}(\beta_2-2\beta_1) \nonumber \\
   &=& -r^2\frac{d}{dr^2}(\ln I_2 - 2\ln I_1)
\end{eqnarray}

Specifying $I$ for quasielectron (quasihole) by
$I^{1qh},~I^{2qh},~I^{1qe},~I^{2qe}$, we can write the following
expression for charge and statistics parameter \cite{KM}
\begin{eqnarray}
  \frac{q^{qh}}{e}&=& \frac{d}{dr^2} \ln I^{1qh} \\
  \frac{q^{qe}}{e}&=& -\frac{d}{dr^2} \ln I^{1qe} \\
  \theta^{qh} &=& -r^2\frac{d}{dr^2}(\ln I^{2qh} - 2\ln I^{1qh}) \\
  \theta^{qe} &=& r^2\frac{d}{dr^2}(\ln I^{2qe} - 2\ln I^{1qe})
\end{eqnarray}

Now from the wavefunctions for
$\psi^{1qh}_{z_0},~\psi^{1qe}_{z_0},~\psi^{2qh}_{z_a,z_b},~\psi^{2qe}_{z_a,z_b}$
given by eqns.(\ref{ee1}),(\ref{ee2}),(\ref{ee3}) and (\ref{e4}), we
have \cite{KM}
\begin{eqnarray}
  I^{1qh}(r^2) &=& \int d^{2N} z ~ \psi_0^2~ |\triangle|^{2m}
    \prod^{N}_{k=1}~|z_k-z_0|^2 \label{iqh}\\
    I^{1qe}(r^2) &=& \int d^{2N} z~  \psi_0^2~ |\triangle|^{2m}
    \prod^{N}_{k=1}\left(|z_k-z_0| ^2-1\right)\label{iqe}\\
    I^{2qh}_{z_a,z_b} &=& \int d^{2N} z ~ \psi_0^2~ |\triangle|^{2m}
    \prod^{N}_{k=1}|(z_k-z_a)(z_k-z_b)|^2\label{iqhh}\\
  I^{2qe}_{z_a,z_b} &=& \int d^{2N} z  ~\psi_0^2~ |\triangle|^{2m}
    \prod^{N}_{k=1}\left(|z_k-z_a|^2
    |z_k-z_b|^2-|2z_k-z_a-z_b|^2 +2\right)\label{iqee}
\end{eqnarray}

From the above results obtained by Kj{\o}nsberg and Myrheim
\cite{KM}, we now consider the case when the quasiparticle is
subjected to the reduced magnetic field $B^*$ after the removal of
the Berry phase associated with even number of flux quanta to the
dynamical phase. It is noted that when the magnetic field $B$ is
reduced to $B^*$, the corresponding magnetic length is changed from
$l$ to $l^*=l\sqrt{m}$ where $m$ is given by $2pn+1$ corresponding
to the filling factor $\nu=\frac{n}{2pn+1}$. Now considering that
the ground state can be thought as the ``vacuum'', we put the
quasiparticle at the position $z_0$ such that it is subjected to the
reduced magnetic field $B^*$. The corresponding change in the value
of the magnetic length $l^*=l\sqrt{m}$, at filling factor
$\nu=\frac{1}{m}=\frac{1}{2p+1}$ will modify the expressions
$I^{qh}_1$ and $I^{qe}_1$ given by eqns.(\ref{iqh}) and (\ref{iqe})
\begin{eqnarray}
    I^{qh}_1&=& \int d^{2N} z ~ \psi_0^2~ |\triangle|^{2m}
    \prod^{N}_{k=1}~m|z_k-z_0|^2 \label{1qh}\\
    I^{qe}_1&=& \int d^{2N} z~  \psi_0^2~ |\triangle|^{2m}
    \prod^{N}_{k=1}\left(m|z_k-z_0| ^2-1\right)\label{1qe}\\
\end{eqnarray}

 Similarly for this quasiparticles, the expressions for $I^{qh}_2$ and $I^{qe}_2$
will be modified as
\begin{eqnarray}
      I^{qh}_{2} &=& \int d^{2N} z ~ \psi_0^2~ |\triangle|^{2m}
    \prod^{N}_{k=1}m^2|(z_k-z_a)(z_k-z_b)|^2\label{2qhh}\\
  I^{qe}_{2} &=& \int d^{2N} z  ~\psi_0^2~ |\triangle|^{2m}
    \prod^{N}_{k=1}~m^2\left(|z_k-z_a|^2
    |z_k-z_b|^2-m|2z_k-z_a-z_b|^2 +2\right)\label{2qee}
\end{eqnarray}

These changes in the expressions due to the change in the
normalization factor for $z$ will modify the numerical results
obtained by Kj{\o}nsberg and Myrheim \cite{KM}. Indeed, the crux of
the problem related to the differences in numerical value for the
charge and statistics parameter for quasihole and quasiparticles  in
ref.[1] lies in the expressions for $I^{qh}_1(I^{qh}_2)$ and
$I^{qe}_1(I^{qe}_2)$. In $I^{qe}_1$ we have the extra factor $(-1)$
associated with the term $|z_k-z_0|^2$. If we could neglect this
factor $(-1)$ then $I^{qh}_1$ and $(I^{qe}_1)$ would have been
identical. However since the average number of electrons within unit
distance from an arbitrary point $z_0$ assuming constant density
within the electron droplet is $\pi/m$, it is very close to $1$ for
$m=3$. But when we consider the effect of the reduced magnetic field
$B^*$ on the quasiparticle, this average number of electrons will
change to $n\pi$ for filling factor $\frac{n}{2pn+1}$ which is
expected to improve the situation.

\section{Numerical Results}
From our above analysis it appears that when the quasiparticle (QH
or QE) is subjected to the reduced magnetic field $B^*$, the
expression for $z_0=r~e^{i\phi}$ in unit of $l^*=l\sqrt{m}$ will
change the parameter $r$ to $\tilde{r}=r/\sqrt{m}$. Similarly, the
parameter related to the distance between two quasiparticles will
also be modified. Taking into account this modification we have
reproduced here the numerical results for the charge $q/e$ and
statistics parameter $\theta$ obtained by Kj{\o}nsberg and Myrheim
\cite{KM} in Figs. 1,2,3 and 4. It is to be remarked that as in the
present formalism a quasiparticle is not a conventional point like
anyon, rather it has an extended structure as certain magnetic flux
lines are attached to it, the size of a quasiparticle can be taken
to be of the order of one magnetic length. In view of this we plot
here the values from $\tilde{r}=1$.

It is found that, as expected the charge and statistics parameter
fits very well with the values $e/3$ and $1/3$ respectively in case
of quasiholes as shown in figs 1 and 3. It may be noted that below
$\tilde{r}<1$, we have observed appreciable deviation from the value
$q=e/3$ and $\theta=1/3$ as has been shown in fig 2 and 6
$(r=\tilde{r}\sqrt{m})$ reported in ref.[1].

However, in case of quasielectrons, we note that the slight
deviation from the exact value for the charge $q=e/3$ is minimal for
different values of $N$ in the range $\tilde{r}=1$ to $3$. Similarly
the average value of the statistics parameter $\theta$ for
quasielectrons becomes close to the value $1/3$ within the range
$\tilde{r}=1$ to $3$ and it deviates largely when $\tilde{r}$
increases. In fact it is expected that we should consider the result
within this range as the edge effect becomes prominent beyond this.
In the perspective of spherical geometry the size of the droplet $R$
is given by $R\sim l\sqrt{N}$. To be consistent with the distance
coordinates $\tilde{r}=r/\sqrt{m}$, we have to consider the droplet
size $\tilde{R}\sim \frac{l}{\sqrt{m}}\sqrt{N}$. So even for
$N=100$, with $m=3$, we note that $\tilde{R}$ is of the order of $5$
units of magnetic length. So it is expected that the edge effect
will be minimal only in the range $\tilde{r}=1$ to $3$. Indeed it
has been observed that all important physics at the edge is not one
dimensional but leads to the formation of striped phase at $\nu=1/3$
caused by long ranged Coulomb interaction \cite{TG}. The amplitude
of the charge density oscillations decays slowly only as a power law
with the distance from the edge. It is found that at $\nu=1/3$ the
quantum Hall edge undergoes a reconstruction as the background
potential softens, whereas quantum Hall edges at higher filling
factors are robust against reconstruction \cite{JNG}. Since the edge
physics at $\nu=1/3$ is not confined to the boundary but extends to
the bulk, we should consider our results within a specific range
\cite{BP}. However, as the size of the droplet varies as
$R\sim\sqrt{N}$, it is expected that the situation will improve for
large $N$.

\begin{figure}
  \centerline{\epsfxsize=8.0in \epsfysize=8.0in \epsfbox{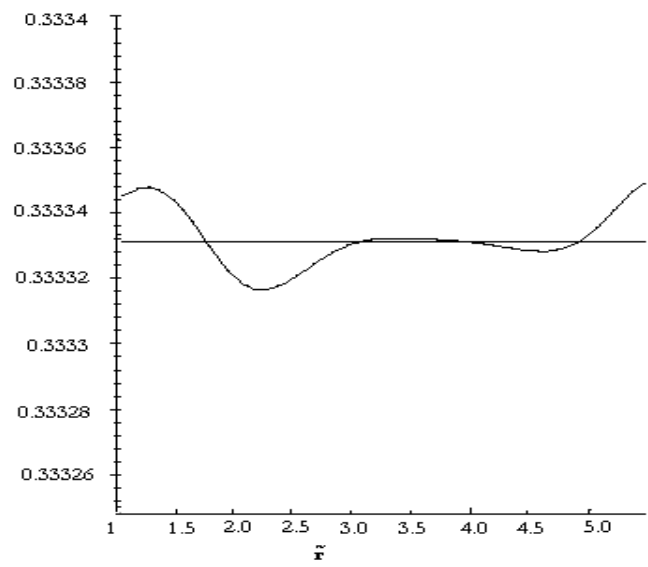}}
  \caption{ The quasihole charge, $q^{qh}/e$, compared to $1/3$, for 100 electrons.}
\label{m1}
\end{figure}

\begin{figure}
  \centerline{\epsfxsize=8.0in \epsfysize=8.0in \epsfbox{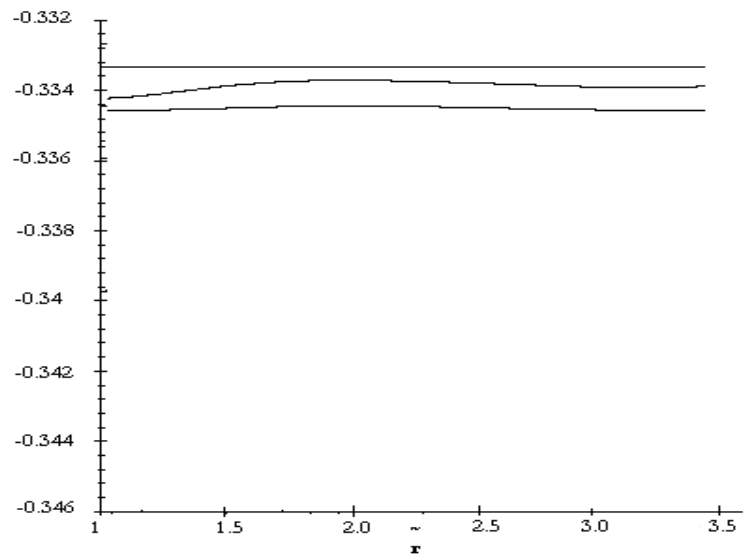}}
  \caption{ Quasielectron charge, $q^{qe}/e$, for 100 and 200 electrons.
  The highest curve is the constant $-1/3$}
\label{m2}
\end{figure}

\begin{figure}
  \centerline{\epsfxsize=8.0in \epsfysize=8.0in \epsfbox{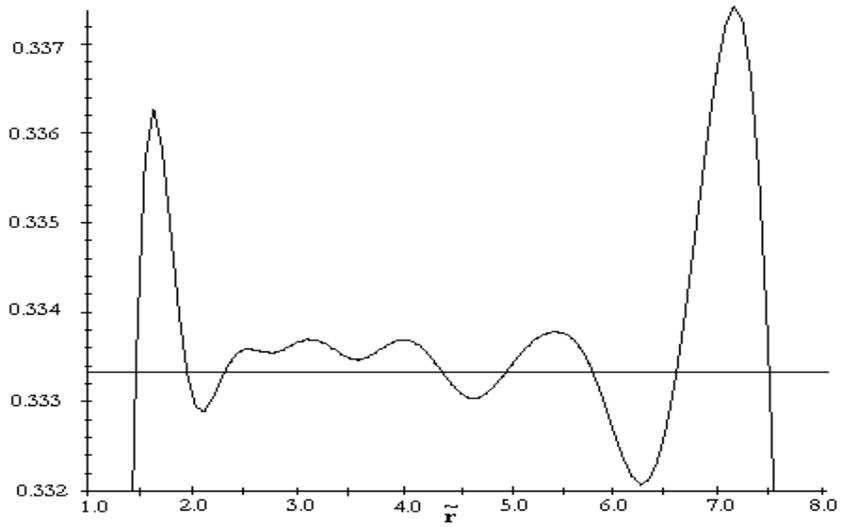}}
  \caption{ Quasihole statistics parameter $\nu^{qh}$ for 100 electrons,
  compared to $1/3$, emphasizing the bulk behavior.}
\label{m3}
\end{figure}

\begin{figure}
  \centerline{\epsfxsize=8.0in \epsfysize=8.0in \epsfbox{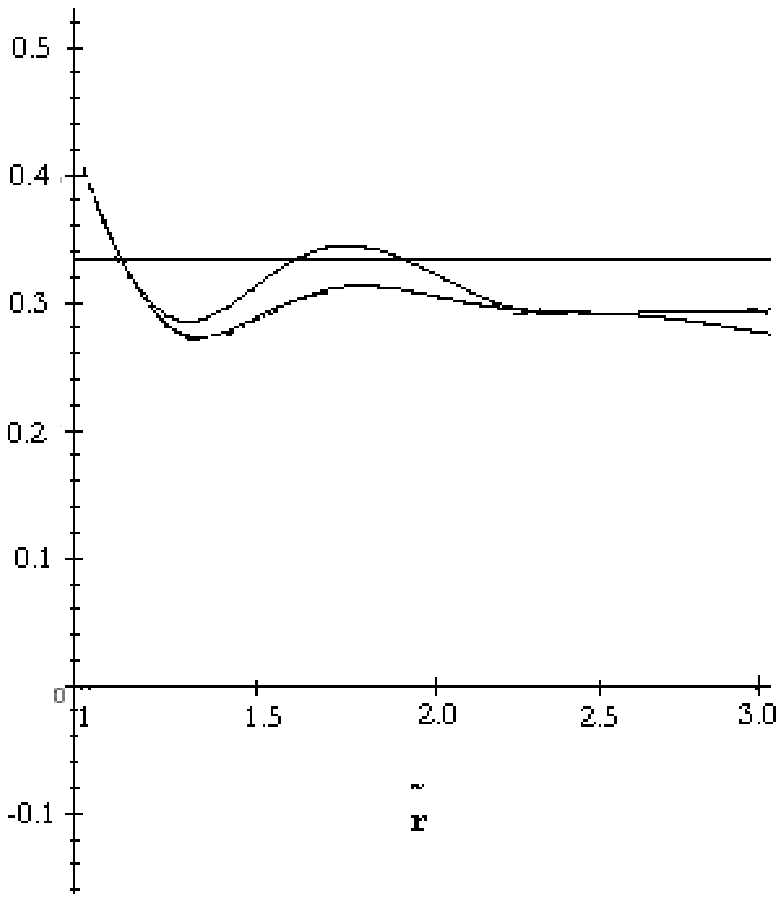}}
  \caption{ Quasielectron statistics parameter $\nu^{qe}$ for 100 and 200
  electrons. The 200 electron curve overshoots the horizontal line $1/3$ at
  $\tilde{r}\approx 1.732$.}
\label{m4}
\end{figure}

In view of these considerations we can argue that the deviation from
the exact value of $e/3$ and $1/3$ for charge and statistics
parameter at $\nu=1/3$ is to be interpreted in terms of the nature
of the quasiparticles and edge physics and it is expected that we
will have exact value in the thermodynamic limit.

\vspace{7 cm}

\section{Discussion}

We have pointed out that when the effect of the removal of the Berry
phase associated with even number of flux quanta to the dynamical
phase is taken into account, we get better estimates of the charge
and statistics parameter of the quasiparticles. Again taking into
account the edge effect in the bulk which is found to be very
prominent at the filling factor $\nu=1/3$, the departure from the
expected value for the charge and statistics parameter for
quasielectrons $q=e/3$ and $\theta=1/3$ can be explained. It may be
recalled that in the composite fermion picture the value of the
statistics parameter $\theta$ is predicted to be $2/3$ for
$\nu=1/3$. Though the numerical computation agrees with the
magnitude, it differs in sign. This raises doubt on the fundamental
interpretation of the CF physics. To cope with this problem it has
been suggested that the insertion of a certain CFQP at a point
perturbs the other CFQP by pushing it to slightly outward from the
loop.

It is to be remarked that  the present formalism does not
incorporate the conventional point anyon picture. Indeed
quasiparticles are particles attached with magnetic flux quanta and
hence should be treated as extended bodies. Fractional charge and
hence fractional statistics is found to be an outcome of Dirac
quantization condition. It may be noted that in this formalism the
very definition of filling factor $\nu$ is associated with the
charge $\nu e$ derived from the Dirac quantization relation. This is
consistent with experimental results and is identical with the
predictions of the CF theory in case of FQH states with $\nu=1/m$
($m$ being an odd integer). But there is a controversy regarding the
charge of quasiparticles in the FQH states with the filling factor
$\nu=n/m$. In the CF model the predicted charge of the
quasiparticles with $\nu=n/m$ is always $1/m$ which is supposed to
be supported by experiments at a bit higher temperature \cite{RPG}
but is in contrast to the experimental result \cite{CH} where it is
shown that the charge of the quasiparticles are $e/3,~2e/5$ and
$3e/7$ at $\nu=1/3,~2/5$ and $3/7$ at extremely low temperature. In
this context we may add that the Dirac quantization condition which
is a consequence of quantum field theory at $T=0$ (no finite
temperature effect is taken into account) our result is expected to
be valid in the close vicinity of $T=0$. However at higher
temperature it may happen that for quasiparticles with $\nu=n/m$,
the system is dissociated into $n$ copies of quasiparticles with
charge $e^*=e/m$ \cite{BBP}.

Finally we may add that as it has been discussed in the previous
section, the edge effect is most prominent at $\nu=1/3$, it is
expected that the numerical experiments will give better result for
other filling factors such as $2/5,~ 3/5,~3/7$ and so on. In a
future study we shall take up this issue.

\end{document}